# Electric field controlled magnetization and transport properties of La$_{0.7}$Ca$_{0.3}$MnO$_3$ ultrathin film

Himanshu Sharma[1,a)], A. Tulapurkar[2], C. V. Tomy[1]

[1]*Department of Physics, Indian Institute of Technology Bombay, Powai, Mumbai – 400 076, India.*

[2]*Department of Electrical Engineering, Indian Institute of Technology Bombay, Powai, Mumbai – 400 076, India.*

We have investigated the effect of electric field control on the magnetization and the transport properties in La$_{0.7}$Ca$_{0.3}$MnO$_3$ (LCMO) ultrathin film (~10 nm) by using it as the semiconductor channel material of a prototypical field effect device and SiO$_2$ as dielectric gate. A large electroresistance (ER) of ~78% at V$_g$ = −8 V is found in LCMO at 200 K. The direct magnetization measurements confirm the formation of a large ferromagnetic phase with the increase in the applied gate voltage at 200 K.

The properties of manganites are sensitive to the intrinsic strain (e.g., doping, structural manipulations, etc)[1] and extrinsic strain (e.g., application of external electric[2] fields or magnetic[3] fields, electromagnetic radiation[4], pressure[5], etc). However, the possibility to modulate the magnetic and transport properties externally in manganite thin films is an idea which drives intensive research in the research areas of spintronics and multiferroics[6]. Over the last decade, properties associated with the external electric field control of manganite thin films (electric field control of magnetic anisotropy, domain structure, spin polarization, critical temperatures[1,8], etc) have attracted a great deal of attention due to the applications envisaged in low-power spintronics[7] and magnetoelectronics[7,8] devices. Even though there have been some reports regarding the electric field control of transport properties in manganites[6-13,16], direct measurements of magnetic properties through magnetization measurements were not reported and the mechanism behind these effects were also not clearly determined.

In this paper, we have investigated the effect of electric field on the magnetization in conjunction with transport properties in La$_{0.7}$Ca$_{0.3}$MnO$_3$ (LCMO) ultrathin film using an insulating-gate in a field effect transistor (FET) structure (Fig. 1). LCMO is used as the semiconductor channel material of a prototypical field effect device, whereas SiO$_2$ is used as an insulating-gate. Application of a gate voltage is expected to induce high electric field as well as strain on the surface of the manganite film. This in turn would change the magnetization and resistivity of the FET channel[9,10,11]. In order to prepare the device (FET structure), we initially deposited a thin film of SiO$_2$ of ~200 nm on an n-type low resistive Si/Pt substrate by the dielectric sputtering method. Subsequently, a layer of LCMO of about 10 nm thickness was deposited on top of this SiO$_2$ thin film using densified (sintered) targets of LCMO. Thin films of LCMO down to 10 nm were grown by a KrF-Pulsed Laser Deposition (PLD) system with λ = 248 nm and energy density[7] of ~2 Jcm$^{-2}$ using the sintered target. The oxygen pressure and substrate temperature during the deposition were 3.5×10$^{-1}$ mbar and 650° C, respectively. After the deposition, the PLD chamber was vented to atmospheric oxygen pressure (~1000 mbar) and then cooled down to room temperature at a rate of 10° C/min[10]. We have shown previously that the ferromagnetic transition temperature (T$_C$) and the saturation magnetization (M$_{sat}$) of LCMO films are sensitive to the film thickness[1]. It was found that the T$_C$ and the M$_{sat}$ values decrease with a decrease in the film thickness below 8 nm. Also, the spin transfer torque effect is proportional to the thickness[17,18]. Thus, a thickness of 10 nm seems to be an optimal choice for the LCMO films for spin-based applications.

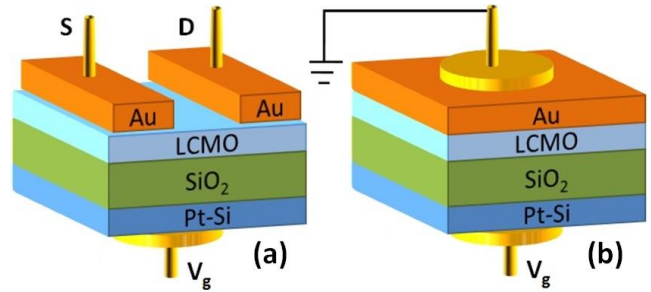

FIG. 1. Schematic of FET device used for (a) magnetization measurements and (b) transport measurements.

In order to fabricate a field effect transistor (FET) structure, Source (S) and Drain (D) channels were made on the manganite film with Au (See Fig. 1(a)) using the optical and e-beam lithography methods. The Si/Pt substrate was used as the gate electrode for transport measurements. However, the top of manganite thin film was completely covered with Au to use it as the top electrode in magnetization measurements, as shown in Fig. 1 (b).

For transport measurements, a cryostat of physical property measurement system (PPMS - Quantum Design Inc., USA) was used for obtaining the magnetic field and the temperature variation. A dual channel Source Meter (Keithley - 2602A) was used for measuring the resistance as well as to apply a gate voltage. The resistance is recorded at different temperatures and with varied gate voltages. In order to ensure that there is no leakage current across the FET structure when a gate a voltage is applied, we have measured the gate current which is found to be almost negligible (< 1 nA). For magnetization measurements, the DC transport option of a SQUID Magnetometer (MPMS-XL, Quantum Design Inc., USA) was used with a suitably modified sample insert rod. The magnetization was recorded as a function of temperature and magnetic field at different applied voltages (electric field) using a Keithley Source Meter (Keithley-2602A).

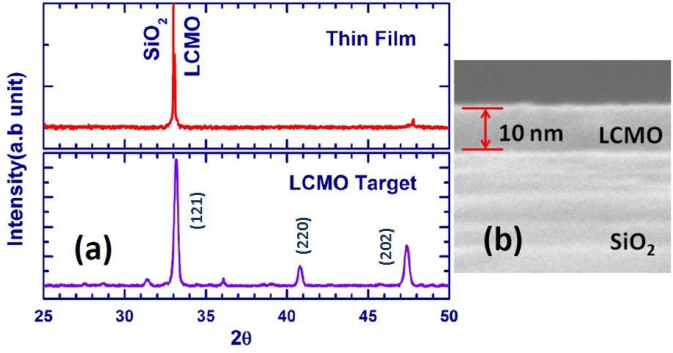

FIG. 2. (a) X-ray diffraction patterns of the LCMO target (bulk) and thin film, (b) Cross section SEM image of LCMO thin film grown on SiO$_2$ thin film.

Figure 2(a) shows the XRD pattern of the sintered LCMO target and the LCMO thin film. From the XRD data (in expanded form), we could clearly identify a strong peak (121) of LCMO which confirms the single phase nature of the film. The cross section SEM image (expanded view) shown in Fig. 2(b) confirms the thickness and uniform growth of the LCMO thin film.

In order to investigate the effect of gate voltage on the transport properties, we first measured the resistance of the FET device as a function of temperature by varying the gate voltage ($V_g$) in the presence of a constant in-plane current of 100 $\mu$A between the source and the drain. Figure 3 shows the variation of the device resistance in applied gate voltages of 0 V, +5V, −3V, −5V and −8V, respectively at a constant applied in-plane magnetic field of 1 kOe. In zero applied gate voltage ($V_g$ = 0 V), the metal to insulator transition (due to double exchange (DE)[6] which leads to the ferromagneitc ordering) is observed close to the magnetic ordering temperature ($T_C$ ~215 K). When a positive gate voltage is applied, we do not observe any appreciable change in the behavior of the resistance variation. In Fig. 3, we have shown one such curve for $V_g$ = +5 V. The metal to insulator transition (MIT) temperature ($T_p$) remains almost the same, even though there is a slight increase in the electro-resistance at the peak value. However, when a negative gate voltage is applied, we observe a drastic decrease in the electroresistance (ER), especially near to the peak temperature ($T_p$), accompanied by a shift of the MIT temperature (by nearly 21% for $V_g$ = −8 V) towards lower temperatures. The maximum change in ER is observed near $T_p$. We have further confirmed this observation by measuring the electroreisistance as a function of gate voltage at 200 K (close to $T_p$), which is shown in Fig. 4.

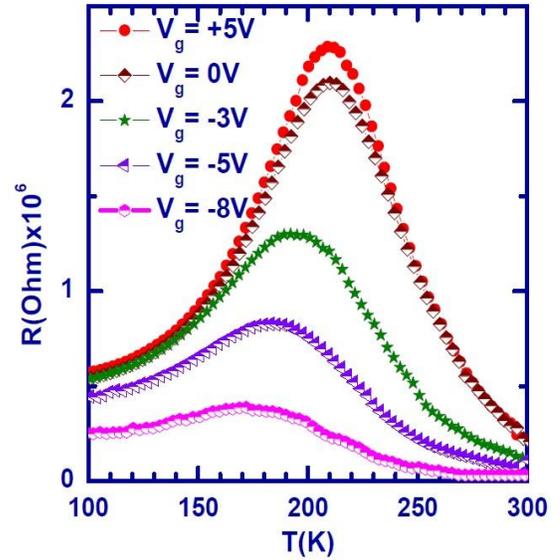

FIG. 3. Temperature dependent resistance of the LCMO channel for applied gate voltages ranging from −8 V to +5 V.

The percentage ER is calculated using the formula, [($R_{V_g}$ − $R_{V_g=0}$)/$R_{V_g=0}$]×100; where $R_{V_g=0}$ and $R_{V_g}$ are the resistances in zero gate voltage and applied gate voltage of $V_g$, respectively. A large electroresistance of ~78% at $V_g$ = −8 V is found in LCMO near $T_P$ (200 K) as shown in Fig. 4. On the other hand, only a small ER of ~ 15 % is observed with the positive applied gate voltage which even saturates for applied gate voltage of > +2 V (see Fig. 4). The value of ER (~78%) for positive gate voltages is comparable to the ER value (76 %) reported by Wu et. al.[9], using a PbZr$_{0.2}$Ti$_{0.8}$O$_3$ (PZT) ferroelectric gate. However, our ER value is much higher than the ER value (~40%) obtained by them when a dielectric STO was used as the gate material in their device. This may be due to the fact that the film used by us is much thinner (~10 nm) than the film of LCMO (50 nm) used by the authors in ref [9].

a) Email: himsharma@iitb.ac.in

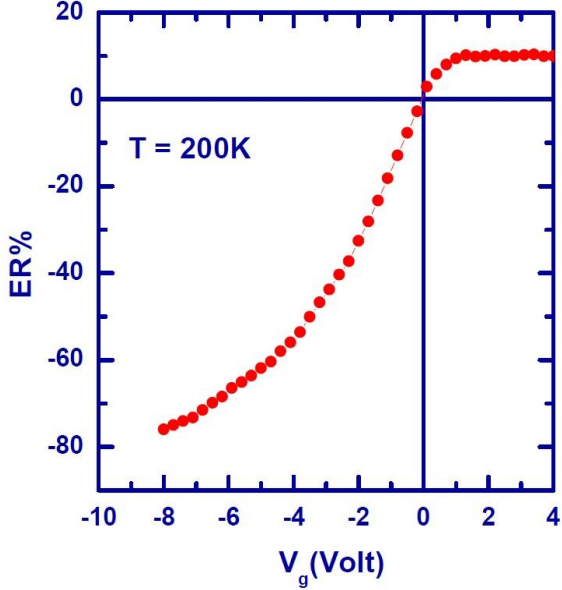

FIG. 4. Electroresistance ER% is plotted against applied gate voltage $V_g$ at 200 K.

Figure 5 shows the magnetoresistance (MR) plots as a function of applied magnetic field at 200 K for different applied gate voltages. The percentage MR is calculated using the formula, $([R(H) - R(0)]/R(0)) \times 100$; where $R(0)$ and $R(H)$ are the resistances in zero magnetic field and applied magnetic field of H, respectively. As we increase the applied negative gate voltage, MR also increases. The value of MR with zero gate voltage is ~38% in a magnetic field of 20 kOe which increases to ~60% when the applied gate voltage is increased to −8 V (see Fig. 5). It is interesting to note that the MR value obtained with $V_g = 0$ V at 20 kOe can even be achieved with a much smaller magnetic field of 8 kOe if a gate voltage of −8 V is applied.

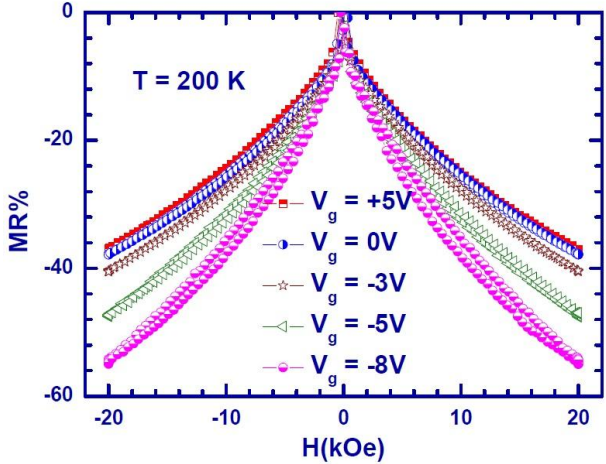

FIG. 5. Magnetoresistance versus applied magnetic field under various applied gate voltages from −8 V to 5 V at 200 K.

It is obvious from the measurements mentioned above, that there is a change in magnetization with the applied gate voltage. Also, in the previous studies[19,20], it is reported in manganites that the observed large change in electroresistance with applied gate voltages (electric field) is possibly associated with modulation of ferromagnetism by applied gate voltage. As the ferromagnetic alignment allows the $e_g$ electrons to hop from $Mn^{+3}$ to $Mn^{+4}$ via the O atom, through the film due to double exchange (DE) and hence the material became metallic[19,20].

In order to investigate this further, we used a direct method by measuring the magnetization by applying different gate voltages (+8 V, 0 V, −5 V and −8 V). The magnetization as a function of temperature of the LCMO channel was recorded with in-plane applied magnetic field of 1 kOe. The LCMO film with zero applied gate voltage exhibits a Curie temperature $T_C$ near 215 K (Fig. 6) which is consistent with the metal to insulator transition seen in the resistance measurements (Fig. 3). As the negative gate voltage increases, a large increase in magnetization with a shift of $T_C$ to lower temperatures is observed which again corroborates the observation from the resistance measurements.

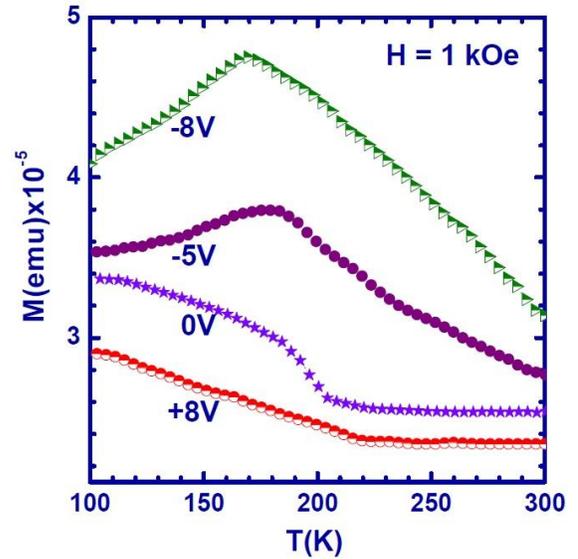

FIG. 6. Temperature dependence of magnetization for the LCMO channel with applied gate voltages of −8 V, −5 V, 0 V and +8 V.

It is to be noted that a decrease in magnetization and a slight increase in $T_C$ is observed with positive applied gate voltages, as shown in Fig. 6 for $V_g = +8$ V. The results observed from magnetization measurements with various gate voltages are consistent with the transport measurements with similar gate voltages, as shown in Fig. 3.

The increase/decrease in magnetization is usually associated with the increase/decrease in the magnetic moments in the system. This is further confirmed through

a)Email: himsharma@iitb.ac.in

the magnetization measurements as a function of magnetic field at 200 K, which is shown in Fig. 7 for applied gate voltages of +8 V, 0 V, −5 V and −8 V, respectively. In zero applied gate voltage, we see a saturation-type magnetization (higher fields may be needed to obtain the full saturation) and coercivity of 50 Oe, consistent with the ferromagnetic nature of this film. For increasing negative gate voltages, we see a clear increase in the saturation magnetization and the coercive field (110 Oe and 180 Oe is for $V_g$ = −5 V and −8 V, respectively) confirming the increase in the ferromagnetic component with negative gate voltage. With $V_g$ = −8 V, we can even get a clear saturation magnetization with field as low as 0.5 kOe. Consistent with the magnetization measurements, we see a decrease in magnetization as well as coercive field for positive applied gate voltages (see Fig. 7 for $V_g$ = +8 V).

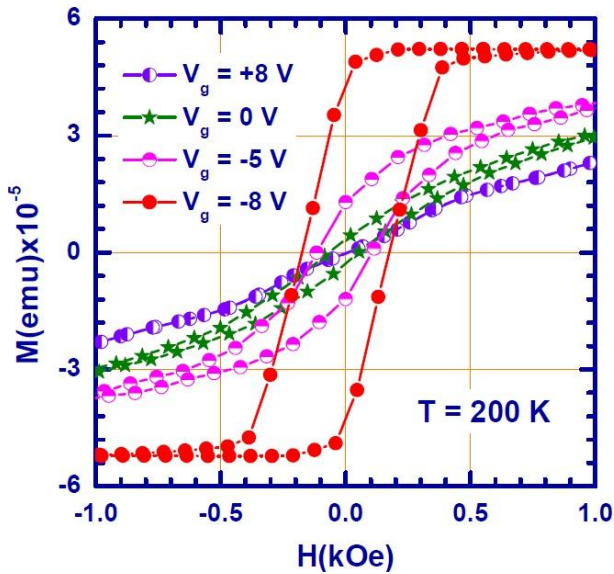

FIG. 7. Magnetic moment versus in-plane applied magnetic field under various applied gate voltages from −8 V to +8 V at 200 K.

We have demonstrated the electric field control of transport properties and magnetization using a field effect transistor (FET) structure with La$_{0.7}$Ca$_{0.3}$MnO$_3$ (LCMO) as the channel material. Our observation of the electric field effect on the transport properties in manganite is in agreement with the observation by Wu. et. al.[9], Pallecchi et. al[19]., and Ogale et. al[20]. We have provided further confirmation of these observations through magnetization measurements in this paper. Even though there is no clear understanding for the role of electric field affecting the double exchange (DE) mechanism, the magnetization data confirms some of the existing explanations in the literature. Pallecchi et. al[19]., have surmised that the electric field modulates the charge carriers, which in turn helps in aligning the magnetic moments[19], whereas, the magnetic field aligns the magnetic moments, which enhances the conductivity. Our observation of the increase in magnetic hysteresis (see in Fig. 7), channel magnetization and conductivity are consistent with the surmise given by Pallecchi et. al.

In conclusion, the effect of electric field on the magnetic properties of LCMO thin films is demonstrated using a FET structure with SiO$_2$ as dielectric gate. The %ER obtained is as comparable to that obtained using a PZT ferroelectric gate. Our magnetization measurements have confirmed the increase in the ferromagnetic phase with negative gate voltages which in turn increases the magnetization and conductivity. The work presented here is a demonstration of a low-power spintronics device (e.g., spintronics field effect transistor) which can be used as low-field switching device. Further, it will be also fascinating to investigate the magnetic anisotropy[15,16] with applied gate voltage using insulating or ferroelectric gate.

We acknowledge the use of MPMS SQUID magnetometer obtained through the DST project (DST Nano Mission: SR/NM/NS-1119/2011). We are grateful to the Department of Electrical Engineering and IITB Nanofabrication facility (IITBNF) for the fabrication of the device.

a)Email: himsharma@iitb.ac.in

[a] Email: himsharma@iitb.ac.in